\documentclass[12pt]{article}
\usepackage{amssymb}
\usepackage{graphicx}
\begin{document}
\newcommand{\beq}{\begin{equation}}
\newcommand{\eeq}{\end{equation}}
\newcommand{\beqa}{\begin{eqnarray}}
\newcommand{\eeqa}{\end{eqnarray}}
\newcommand{\beqar}{\begin{eqnarray*}}
\newcommand{\eeqar}{\end{eqnarray*}}
\newcommand{\al}{\alpha}
\newcommand{\be}{\beta}
\newcommand{\del}{\delta}
\newcommand{\D}{\Delta}
\newcommand{\eps}{\epsilon}
\newcommand{\ga}{\gamma}
\newcommand{\Ga}{\Gamma}
\newcommand{\ka}{\kappa}
\newcommand{\nn}{\nonumber}
\newcommand{\inn}{\!\cdot\!}
\newcommand{\h}{\eta}
\newcommand{\ii}{\iota}
\newcommand{\kk}{\varphi}
\newcommand\F{{}_3F_2}
\newcommand{\la}{\lambda}
\newcommand{\La}{\Lambda}
\newcommand{\na}{\prt}
\newcommand{\Om}{\Omega}
\newcommand{\om}{\omega}
\newcommand{\p}{\Phi}
\newcommand{\sig}{\sigma}
\renewcommand{\t}{\theta}
\newcommand{\z}{\zeta}
\newcommand{\ssc}{\scriptscriptstyle}
\newcommand{\eg}{{\it e.g.,}\ }
\newcommand{\ie}{{\it i.e.,}\ }
\newcommand{\labell}[1]{\label{#1}} 
\newcommand{\reef}[1]{(\ref{#1})}
\newcommand\prt{\partial}
\newcommand\veps{\varepsilon}
\newcommand{\pol}{\varepsilon}
\newcommand\vp{\varphi}
\newcommand\ls{\ell_s}
\newcommand\cF{{\cal F}}
\newcommand\cA{{\cal A}}
\newcommand\cS{{\cal S}}
\newcommand\cT{{\cal T}}
\newcommand\cV{{\cal V}}
\newcommand\cL{{\cal L}}
\newcommand\cM{{\cal M}}
\newcommand\cN{{\cal N}}
\newcommand\cG{{\cal G}}
\newcommand\cK{{\cal K}}
\newcommand\cH{{\cal H}}
\newcommand\cI{{\cal I}}
\newcommand\cJ{{\cal J}}
\newcommand\cl{{\iota}}
\newcommand\cP{{\cal P}}
\newcommand\cQ{{\cal Q}}
\newcommand\cg{{\it g}}
\newcommand\cR{{\cal R}}
\newcommand\cB{{\cal B}}
\newcommand\cO{{\cal O}}
\newcommand\tcO{{\tilde {{\cal O}}}}
\newcommand\bz{\bar{z}}
\newcommand\bb{\bar{b}}
\newcommand\ba{\bar{a}}
\newcommand\bg{\bar{g}}
\newcommand\bc{\bar{c}}
\newcommand\bomega{\bar{\omega}}
\newcommand\bH{\bar{H}}
\newcommand\bw{\bar{w}}
\newcommand\bX{\bar{X}}
\newcommand\bK{\bar{K}}
\newcommand\bA{\bar{A}}
\newcommand\bZ{\bar{Z}}
\newcommand\bxi{\bar{\xi}}
\newcommand\bphi{\bar{\phi}}
\newcommand\bpsi{\bar{\psi}}
\newcommand\bprt{\bar{\prt}}
\newcommand\bet{\bar{\eta}}
\newcommand\btau{\bar{\tau}}
\newcommand\hF{\hat{F}}
\newcommand\hA{\hat{A}}
\newcommand\hT{\hat{T}}
\newcommand\htau{\hat{\tau}}
\newcommand\hD{\hat{D}}
\newcommand\hf{\hat{f}}
\newcommand\hK{\hat{K}}
\newcommand\hg{\hat{g}}
\newcommand\hp{\hat{\Phi}}
\newcommand\hi{\hat{i}}
\newcommand\ha{\hat{a}}
\newcommand\hb{\hat{b}}
\newcommand\hQ{\hat{Q}}
\newcommand\hP{\hat{\Phi}}
\newcommand\hS{\hat{S}}
\newcommand\hX{\hat{X}}
\newcommand\tL{\tilde{\cal L}}
\newcommand\hL{\hat{\cal L}}
\newcommand\MZ{\mathbb{Z}}
\newcommand\tG{{\tilde G}}
\newcommand\tg{{\tilde g}}
\newcommand\tphi{{\widetilde \Phi}}
\newcommand\tPhi{{\widetilde \Phi}}
\newcommand\ti{{\tilde i}}
\newcommand\tj{{\tilde j}}
\newcommand\tk{{\tilde k}}
\newcommand\tl{{\tilde l}}
\newcommand\ttm{{\tilde m}}
\newcommand\tn{{\tilde n}}
\newcommand\ta{{\tilde a}}
\newcommand\tb{{\tilde b}}
\newcommand\tc{{\tilde c}}
\newcommand\td{{\tilde d}}
\newcommand\tm{{\tilde m}}
\newcommand\tmu{{\tilde \mu}}
\newcommand\tnu{{\tilde \nu}}
\newcommand\talpha{{\tilde \alpha}}
\newcommand\tbeta{{\tilde \beta}}
\newcommand\trho{{\tilde \rho}}
 \newcommand\tR{{\tilde R}}
\newcommand\teta{{\tilde \eta}}
\newcommand\tF{{\widetilde F}}
\newcommand\tK{{\widetilde K}}
\newcommand\tE{{\tilde E}}
\newcommand\tpsi{{\tilde \psi}}
\newcommand\tX{{\widetilde X}}
\newcommand\tD{{\widetilde D}}
\newcommand\tO{{\widetilde O}}
\newcommand\tS{{\tilde S}}
\newcommand\tB{{\tilde B}}
\newcommand\tA{{\widetilde A}}
\newcommand\tT{{\widetilde T}}
\newcommand\tC{{\widetilde C}}
\newcommand\tV{{\widetilde V}}
\newcommand\thF{{\widetilde {\hat {F}}}}
\newcommand\Tr{{\rm Tr}}
\newcommand\tr{{\rm tr}}
\newcommand\STr{{\rm STr}}
\newcommand\hR{\hat{R}}
\newcommand\M[2]{M^{#1}{}_{#2}}

\newcommand\bS{\textbf{ S}}
\newcommand\bI{\textbf{ I}}
\newcommand\bJ{\textbf{ J}}

\begin{titlepage}
\begin{center}

\vskip 2 cm
{\LARGE \bf  The data on the boundary     \\ \vskip .2 cm   at order $\alpha'$
 }\\
\vskip 1.25 cm
  Mohammad R. Garousi\footnote{garousi@um.ac.ir}

\vskip 1 cm
{{\it Department of Physics, Faculty of Science, Ferdowsi University of Mashhad\\}{\it P.O. Box 1436, Mashhad, Iran}\\}
\vskip .1 cm
 \end{center}

\begin{abstract}

The least action principle indicates that for the open spacetime manifolds, there are data on the boundary.  Recently, it has been proposed that the data for the effective actions at order $\alpha'$ are the values of the massless  fields and their first derivatives.  These data  should be respected by  the  T-duality transformations  at order $\alpha'$. Moreover, the  T-duality  transformations should not change the unit vector to the boundary which in turns implies that  the base space metric should be also invariant. 
Assuming such restricted T-duality transformations, we show that the  transformation of the circular reduction of the parity-odd part of the effective action of the heterotic string theory at order $\alpha'$ under the Buscher rules  is cancelled  by some  total derivative terms  and by some restricted T-duality transformations  at order $\alpha'$. Using the Stokes' theorem, we then show that the boundary terms in the base space corresponding to the total derivative terms are  $exactly$ cancelled  by transformation of  the circular reduction of the  Gibbons-Hawking boundary term under the above restricted T-duality transformations. These calculations confirm the above proposal for the data on the boundary for the effective actions at order $\alpha'$.
\end{abstract}

\end{titlepage}

\section{Introduction}

The least action principle indicates  that there are data on the boundary in the string field theory.   As it has been argued in \cite{Myers:1987yn}, in the string field theory, the data  are  the values of   the string field  on the boundary.   The string field has  massless  fields and infinite tower of  massive fields. When integrating out the massive fields to produce  the effective actions which involve only the massless fields and their derivatives, the data on the boundary should  be rearranged  as the values of the massless fields and their derivatives. It has been proposed in \cite{Garousi:2021cfc} that for the effective action at the leading order of  $\alpha'$,  the data are  only the values of the massless fields. 
For the effective action at order $\alpha'$,  the data are  the values of the massless fields and their first derivatives.  For the effective action at order $\alpha'^n$, the data are  the values of the massless fields and their  derivatives up to order $n$.
It has been shown  in \cite{Garousi:2021yyd} that for the open spacetime manifolds, the higher-derivative field redefinitions should be restricted to those which respect the above data on the boundary.  We propose that  the global symmetries of the classical effective actions should also respect the above data on the boundary. 

It is known that   the   Kaluza-Klein (KK) reduction  of the classical  effective actions of  the bosonic  and the heterotic string theories on torus $T^d$  are  invariant under the rigid $O(d,d)$-transformations  at all orders of $\alpha'$ \cite{Sen:1991zi,Hohm:2014sxa}.  It is   speculated in  \cite{Garousi:2022ovo} that the   effective actions of string theory at the critical dimension are independent of the spacetime  manifolds. Hence, if one uses the particular closed spacetime manifold which  includes the  compact sub-manifold $T^d$ and uses the KK reduction, then the non-geometrical subgroup of the $O(d,d)$-group  may be used to interconnect the coefficients of the original  bulk couplings.  
 This idea has been used in \cite{Garousi:2019mca,Garousi:2020gio} for the circular reduction to find all bulk couplings of dilaton, $B$-field and metric at orders $\alpha'^2,\alpha'^3$ up to overall factors.  The background independence also indicates that  the global $O(d,d)$-symmetry should be  the symmetry   of the more general open spacetime manifolds that have boundary. The non-geometrical subgroup in this case may also  connect the coefficients of the bulk couplings to the coefficients of the boundary couplings. This idea has been used in \cite{Garousi:2019xlf} for the circular reduction to reproduce the Gibbons-Hawking boundary term \cite{Gibbons:1976ue} and used in \cite{Garousi:2021cfc,Garousi:2021yyd}  to find the boundary couplings at orders $\alpha'$ in the bosonic string theory.  
 
 The  T-duality transformations in the non-geometrical subgroup of the $O(d,d)$-group  have $\alpha'$-expansion in both closed and open  spacetime manifolds.  In the open spacetime manifolds,  we propose that the  transformations   should not change the data on the boundary.  This has effect on both the data on the boundary and on the T-duality transformations. The $\alpha'$-expansion of the T-duality   dictates that  the data on the boundary should also have an $\alpha'$-expansion,    \ie  the T-duality transformations that  have  $\alpha'$-expansion, can not be consistent with the boundary data  in which  only the values of the massless fields are known. The proposal  also   produces a constraint on the T-duality  transformations.  To see this point, we note that the  T-duality transformations at order $\alpha'$ are  applied on the leading order effective action in which only the values of the massless fields are known, to produce couplings for the effective action at order $\alpha'$ in which, according to \cite{Garousi:2021cfc},  the values of the massless fields and their first derivatives are known.  Hence, the transformations at order $\alpha'$ should involve only the massless fields and their first derivatives.  The  T-duality transformations at order $\alpha'^2$ are  applied on the leading order effective action to produce couplings for the effective action at order $\alpha'^2$ in which the values of the massless fields and their first and  second derivatives are known.  Hence, the transformations at order $\alpha'^2$ should involve only the massless fields and their first and second derivatives. Similarly for the transformations at the higher orders of $\alpha'$. 

On the other hand, in the presence of boundary, there is a unit vector  orthogonal to the boundary that should be inert under the T-duality transformations at any order of $\alpha'$. Moreover, in order that the length of the vector remains fixed, the metric should also be invariant under the T-duality  transformations at any order of $\alpha'$.   Hence,  apart from the unit vector and the metric which are invariant, the restricted non-geometrical transformations at the leading order of $\alpha'$ should involve only the massless fields, and  at order $\alpha'$ they should involve only the massless fields and their first derivatives. Similarly for the higher orders of $\alpha'$.

Using the circular reduction, it has been shown in \cite{Garousi:2019xlf} that the invariance of the leading order effective action under the  non-geometrical $\MZ_2 $-subgroup of  the rigid $O(1,1)$-group, produces the following standard effective action up to the overall factor:
\beqa
\bS^{(0)}\!+\!\prt\!\!\bS^{(0)}
=-\frac{2}{\kappa^2}\Bigg[\!\int d^{D}x \sqrt{-G} e^{-2\Phi} \left(\!  R + 4\nabla_{\mu}\Phi \nabla^{\mu}\Phi-\frac{1}{12}H^2\right)+ 2\int d^{D-1}\sigma\sqrt{| g|}  e^{-2\Phi}K\Bigg]\labell{baction}
\eeqa
where $\kappa$ is related to the $D$-dimensional Newton's constant and   the last term is the Gibbons-Hawking boundary term \cite{Gibbons:1976ue}. In this term,  $g$ is the determinant of the induced metric. The $\MZ_2 $-transformations in this case are the Buscher rules \cite{Buscher:1987sk} which involve no derivative of the base space fields, \ie
\beqa
&&\varphi'= -\varphi
\,\,\,,\,\,g'_{a }= b_{a }\,\,\,,\,\, b'_{a }= g_{a } \,\,\,,\,\,\nn\\
&&\bg_{ab}'=\bg_{ab} \,\,\,,\,\,\bb_{ab}'=\bb_{ab} \,\,\,,\,\,  \bar{\phi}'= \bar{\phi}\,\,\,,\,\, n_a'=n_a\labell{T220}
\eeqa
where the base space fields are defined in the following KK reduction:
\beqa
&&G_{\mu\nu}=\left(\matrix{\bg_{ab}+e^{\varphi}g_{a }g_{b }& e^{\varphi}g_{a }&\cr e^{\varphi}g_{b }&e^{\varphi}&}\right),B_{\mu\nu}= \left(\matrix{\bb_{ab}+\frac{1}{2}b_{a }g_{b }- \frac{1}{2}b_{b }g_{a }&b_{a }\cr - b_{b }&0&}\right),\nn\\&& \Phi=\bar{\phi}+\varphi/4\,,\quad n^{\mu}=(n^a,0)\labell{reduc}
\eeqa
Note that the base space unit vector $n^a$ and metric $\bg_{ab}$ are invariant.  The data  in this case are the values of the massless field   on the boundary.

Using the circular reduction and the cosmological reduction, it has been shown in \cite{Garousi:2021yyd} that the invariance of the effective action  under the  T-duality groups $O(1,1)$ and $O(d,d)$, respectively,  can produce  the following even-parity  bulk and boundary couplings at order $\alpha'$:
\beqa
\bS^{(1)}
&\!\!\!\!=\!\!\!\!\!&-\frac{48a_1}{\kappa^2} \int_M d^{D}x \sqrt{-G} e^{-2\Phi}\Bigg[R^2_{\rm GB}+\frac{1}{24} H_{\alpha }{}^{\delta \epsilon } H^{\alpha \beta
\gamma } H_{\beta \delta }{}^{\varepsilon } H_{\gamma \epsilon
\varepsilon }-\frac{1}{8}  H_{\alpha \beta }{}^{\delta }
H^{\alpha \beta \gamma } H_{\gamma }{}^{\epsilon \varepsilon }
H_{\delta \epsilon \varepsilon }\nn\\&&\qquad\qquad +
R^{\alpha \beta }H_{\alpha }{}^{\gamma \delta } H_{\beta \gamma \delta }  -\frac{1}{12} R H_{\alpha
\beta \gamma } H^{\alpha \beta \gamma }  -\frac{1}{2} H_{\alpha }{}^{\delta \epsilon } H^{
\alpha \beta \gamma } R_{\beta \gamma \delta \epsilon
}\nn\\&&\qquad\qquad +4R \nabla_{\alpha }\Phi
\nabla^{\alpha }\Phi -16
 R^{\alpha \beta }\nabla_{\alpha }\Phi \nabla_{\beta
}\Phi \Bigg]\labell{ffinal}\\
\prt\!\!\bS^{(1)}&\!\!\!\!=\!\!\!\!\!&-\frac{48a_1}{\kappa^2}\int d^{D-1}\sigma\sqrt{|g|}  e^{-2\Phi}\Bigg[Q_2+ \frac{4}{3}n^2 n^{\alpha }
n^{\beta } \nabla_{\gamma }\nabla^{\gamma }K_{\alpha \beta
}-\frac{1}{6} H_{\beta \gamma \delta } H^{\beta \gamma \delta
} K^{\alpha }{}_{\alpha } +  H_{\alpha }{}^{\gamma \delta }
H_{\beta \gamma \delta } K^{\alpha \beta } \nn\\&&
\qquad\quad +  n^2H_{\alpha }{}^{\delta \epsilon } H_{\beta
\delta \epsilon } K^{\gamma }{}_{\gamma } n^{\alpha }
n^{\beta }  - 2 n^2H_{\beta
}{}^{\delta \epsilon } H_{\gamma \delta \epsilon } n^{\alpha }
n^{\beta } n^{\gamma } \nabla_{\alpha }\Phi + 8 K^{\beta
}{}_{\beta } \nabla_{\alpha }\Phi \nabla^{\alpha }\Phi\labell{fff}\\&&\qquad\quad - 16
n^2K^{\gamma }{}_{\gamma } n^{\alpha } n^{\beta }
\nabla_{\alpha }\Phi \nabla_{\beta }\Phi - 16 K_{\alpha
\beta } \nabla^{\alpha }\Phi \nabla^{\beta }\Phi + \frac{32}{3}
n^2n^{\alpha } n^{\beta } n^{\gamma } \nabla_{\alpha }\Phi
\nabla_{\beta }\Phi \nabla_{\gamma }\Phi \Bigg]\nn
\eeqa
where $n^2=n^\mu n_\mu$, $R^2_{\rm GB}$ is the Gauss-Bonnet gravity couplings   and $Q_2$ is the Chern-Simons boundary couplings.
The  T-duality  $\MZ_2$-transformation   in this case are
\beqa
&&\varphi'= -\varphi+\alpha'\Delta\vp
\,\,\,,\,\,g'_{a }= b_{a }+\alpha'e^{\vp/2}\Delta g_a\,\,\,,\,\, b'_{a }= g_{a }+\alpha'e^{-\vp/2}\Delta b_a \,\,\,,\,\,\nn\\
&&\bg_{ab}'=\bg_{ab} \,\,\,,\,\,\bH_{abc}'=\bH_{abc}+\alpha'\Delta\bH_{abc} \,\,\,,\,\,  \bar{\phi}'= \bar{\phi}+\alpha'\Delta\bphi\,\,\,,\,\, n_a'=n_a\labell{T22}
\eeqa
As in the leading order,  the base space unit vector $n^a$ and metric $\bg_{ab}$ are invariant. In the above equation, $\bH$  which is defined as $\bH_{abc}=3\prt_{[a}\hat{b}_{bc]}-3g_{[a}W_{bc]}$ where $\hat{b}_{ab}=\bb_{ab}+\frac{1}{2}b_{a }g_{b }- \frac{1}{2}b_{b }g_{a }$,  is the  torsion in the base space. 
The deformations in \reef{T22} corresponding to the $\alpha'$-order actions  \reef{ffinal}, \reef{fff}  involve only the massless fields and their first derivatives which are consistent with the proposed data on the boundary \cite{Garousi:2021cfc} in which the values of the massless fields and their first derivative are known. They are  \cite{Garousi:2021yyd}
\beqa
\Delta\bphi&=&3a_1 e^{\vp} V_{ab} V^{ab} -3a_1 e^{-\vp}W_{ab} W^{ab} +48a_1\nabla_{a}\vp \nabla^{a}\bphi \labell{dbH2}\\
  \Delta\vp&=& 24 a_{1} e^{\vp} V_ {ab} V^{ab} + 24 a_{1}  e^{-\vp}W_{ab}
W^{ab} + 48 a_{1} \nabla_{a}\vp \nabla^{a}\vp \nn\\
  \Delta g_{a}&=& 24 a_{1}e^{ \vp/2}\bH_{abc} V^{bc}-96 a_1e^{-\vp/2}
W_{ab} \nabla^{b}\bphi +24 a_1e^{ -\vp/2} W_{ab} \nabla^{b}\vp\nn\\
   \Delta b_{a}&=&- 24 a_{1}e^{- \vp/2}\bH_{abc} W^{bc} +96 a_1e^{\vp/2}
V_{ab} \nabla^{b}\bphi +24 a_1e^{ \vp/2} V_{ab} \nabla^{b}\vp\nn\\
   \Delta\bH_{abc}&=&-144 a_{1}^{}\prt_{[a}(W_{b}{}^{d} V_{c]d} )+144 a_1 \prt_{[a}(\bH_{bc]d}
\nabla^{d}\vp)-3e^{\vp/2} V_{[ab}\Delta g_{c]}-3e^{-\vp/2} W_{[ab}\Delta b_{c]}
\nn
\eeqa
In the above  equation, $V_{ab}$ is the field strength of the $U(1)$ gauge field $g_{a}$, \ie $V_{ab}=\prt_{a}g_{b}-\prt_{b}g_{a}$, and $W_{\mu\nu}$ is the  field strength of the $U(1)$ gauge field $b_{a}$, \ie $W_{ab}=\prt_{a}b_{\nu}-\prt_{b}b_{a}$. Note that even though the transformation of the torsion involves the first and  the  second derivative terms, however, the transformation of the base space field $\bb_{ab}$  has only the first  derivative terms.  The overall factor for the bosonic string theory  is $a_1=\alpha'/96$ and for the heterotic theory  is $a_1=\alpha'/192$. Using the restricted field redefinitions, it has been shown in \cite{Garousi:2021yyd} that the bulk action \reef{ffinal}  is the same as the  Meissner action \cite{Meissner:1996sa} and the boundary action \reef{fff} is the same as the  boundary action corresponding to   the  Meissner action that has been found in \cite{Garousi:2021cfc}.  

The heterotic theory has another bulk coupling at order $\alpha'$ which is odd under the parity. In this paper,  by studying the invariance of this term under the global $O(1,1)$-transformations, we are going to confirm   the proposal that the $\MZ_2$-transformations at order $\alpha'$ can not include the second derivatives of the base space fields. This in turn confirms that the data on the boundary for the effective action at order $\alpha'$ are values of  the massless field and their first derivatives. In the next section we are going to show that the parity odd term of the heterotic effective action at order $\alpha'$ is invariant under the deformed Buscher rules which involve only the first derivatives of the massless fields. In section 3, we briefly discuss our results and show that if the non-geometrical transformations include  the  second derivatives of the massless fields, then it breaks the data on the boundary and hence the $O(1,1)$-symmetry is broken by the boundary. 

\section{$O(1,1)$-symmetry of odd-parity coupling at order $\alpha'$}

The heterotic string theory has anomaly which can be cancelled by assuming the gauge group to be  $SO(32)$ and the $B$-field to have  the non-standard gauge transformations and local Lorentz-transformations \cite{Green:1984sg}.  For zero gauge field that we consider in this paper,  the non-standard local Lorentz-transformation for the $B$-field is 
\beqa
B_{\mu\nu}\rightarrow B_{\mu\nu}+\alpha' \prt_{[\mu}\Lambda_i{}^j\omega_{\nu]j}{}^i\labell{deform}
\eeqa
where $\Lambda_i{}^j$ is the matrix of the   Lorentz-transformations and $\omega_{\mu i}{}^j$ is spin connection. The invariance under the above local Lorentz-transformations then requires the  $B$-field field strength in \reef{baction}, \reef{ffinal}, \reef{fff} to be replaced by new field strength that is invariant under the above transformation, \ie
\beqa
H_{\mu\nu\alpha}&\rightarrow H_{\mu\nu\alpha}+\Omega_{\mu\nu\alpha}\labell{replace}
\eeqa
where the  Chern-Simons three-form $\Omega$   is 
\beqa
\Omega_{\mu\nu\alpha}&=&\omega_{[\mu i}{}^j\prt_\nu\omega_{\alpha] j}{}^i+\frac{2}{3}\omega_{[\mu i}{}^j\omega_{\nu j}{}^k\omega_{\alpha]k}{}^i\,\,;\,\,\,\omega_{\mu i}{}^j=\prt_\mu e_\nu{}^j e^\nu{}_i-\Gamma_{\mu\nu}{}^\rho e_\alpha{}^j e^\nu{}_i
\eeqa
 where $e_\mu{}^i e_\nu{}^j\eta_{ij}=G_{\mu\nu}$. The spin connection with subscript  indices $\omega_{\mu\nu\alpha}=e_\nu{}^i e_\alpha{}^j\omega_{\mu i j}$ is antisymmetric with respect to its last two indices. The replacement \reef{replace} into \reef{baction}, produces no boundary coupling and produces the following  bulk term  at order $\alpha'$: 
\beqa
\bS^{(1)}_O&=&-\frac{2\alpha'}{\kappa^2}\int d^{10} x \sqrt{-G} \,e^{-2\Phi}\left(-\frac{1}{6}H_{\mu\nu\alpha}\Omega^{\mu\nu\alpha}\right)\labell{CS}
\eeqa
which is odd under the parity. We are going to study in details the invariance of the above term under the $O(1,1)$-transformations after using the circular reduction.

The KK reduction of the frame $e_\mu{}^i$ is  
\beqa
&&e_\mu{}^i=\left(\matrix{\bar{e}_a{}^\ti & 0 &\cr e^{\varphi/2}g_{a }&e^{\varphi/2}&}\right)
\eeqa
where  $\bar{e}_a{}^\ti \bar{e}_b{}^\tj\eta_{\ti\tj}=\bg_{ab}$. The above reduction is  consistent with the KK reduction of metric in \reef{reduc}. Using this reduction and the reductions in \reef{reduc}, one finds the circular reduction of the action \reef{CS} has three and four flux terms. They are \footnote{We have used the package "xAct"  \cite{Nutma:2013zea} for performing the calculations in this paper.} 
\beqa
S^{(1)}_O&\!\!\!\!\!=\!\!\!\!\!&-\frac{2\alpha'}{\kappa'^2}\int d^{9} x \sqrt{-\bg} \,e^{-2\bphi}\Big[\frac{1}{24} e^{\vp} V_{a}{}^{c} V^{ab} V_{b}{}^{d} W_{cd} + 
\frac{1}{48} e^{\vp} V_{ab} V^{ab} V^{cd} W_{cd} -  
\frac{1}{6} V^{ab} W^{cd} \bomega_{ca}{}^{e} \bomega_{dbe}\nn\\&& -  
\frac{1}{24} e^{\vp} \bH_{cde} V^{ab} V^{cd} \bomega^{e}{}_{ab} + \frac{2}{9} \bH_{adf} \bomega^{abc} \bomega^{d}{}_{b}{}^{e} \bomega^{f}{}_{ce} + \frac{1}{12} e^{\vp} 
\bH_{bcd} V^{ab} \nabla_{a}V^{cd}  -  
\frac{1}{6} \bH_{ade} \bomega^{abc} \nabla^{e}\bomega^{d}{}_{bc}\nn\\&&+ \frac{1}{24} e^{\vp} 
\bH_{bcd} V_{a}{}^{b} V^{cd} \nabla^{a}\vp + \frac{1}{12} 
V^{bc} W_{a}{}^{d} \bomega_{dbc} \nabla^{a}\vp + 
\frac{1}{24} V^{bc} W_{bc} \nabla_{a}\vp \nabla^{a}\vp -  
\frac{1}{12} W^{ab} \bomega_{a}{}^{cd} \nabla_{b}V_{cd}\nn\\&& -  
\frac{1}{12} W^{bc} \nabla^{a}\vp \nabla_{c}V_{ab} -  
\frac{1}{12} V^{ab} W_{a}{}^{c} \nabla_{c}\nabla_{b}\vp + 
\frac{1}{12} V^{ab} W^{cd} \nabla_{d}\bomega_{cab} + 
\frac{1}{12} e^{\vp} \bH_{bcd} V^{ab} \nabla^{d}V_{a}{}^{c}\Big]\labell{rCS}
\eeqa
where $\kappa'$ is related to the 9-dimensional Newton's constant and  $\bomega_{abc}$ is the base space spin connection.    Note that the above reduction is covariant in the base space and is invariant under the $U(1)\times U(1)$ gauge transformations. However,  as its  original action \reef{CS} which is not invariant under the local Lorentz-transformations, the above action  is not invariant under the base space local Lorentz-transformations either. Using the fundamental requirement that the frame $\bar{e}_a{}^\ti$ is covariantly constant, \ie $\nabla_{a}\bar{e}_b{}^\ti=0$, one may rewrite the above expression in terms of the flat space fluxes $\bH_{\ti\tj\tk},\bomega_{\ti\tj\tk},V_{\ti\tj},W_{\ti\tj},\nabla_\ti\vp$ and their flat derivatives, \eg
\beqa
\nabla_a V_{bc}&=&\bar{e}_a{}^\ti \bar{e}_b{}^\tj \bar{e}_c{}^\tk  (D_\ti V_{\tj\tk}+\bomega_{\ti\tj}{}^\ttm V_{\ttm\tk}+\bomega_{\ti\tk}{}^\ttm V_{\tj\ttm})
\eeqa
where the flat derivative is $D_\ti=e^a{}_\ti\prt_a$. In either flat space or curved space fluxes, the reduction \reef{rCS} involves only three and four fluxes.   We continue our calculations in this paper with the curved  space tensors.

The transformation of \reef{rCS} under the Buscher rules \reef{T220} becomes
\beqa
\delta S_O^{(1)}&\equiv & S_O^{(1)}-S_O^{(1)'}\,=\,-\frac{2\alpha'}{\kappa'^2}\int d^{9} x \sqrt{-\bg} \,e^{-2\bphi}\Big[\frac{1}{24} e^{\vp} V_{a}{}^{c} V^{ab} V_{b}{}^{d} W_{cd} + 
\frac{1}{48} e^{\vp} V_{ab} V^{ab} V^{cd} W_{cd}\nn\\&& -  
\frac{1}{24} e^{-\vp}V^{ab} W_{a}{}^{c} W_{b}{}^{d} W_{cd} -  
\frac{1}{48}e^{-\vp}V^{ab} W_{ab} W_{cd} W^{cd}  + \frac{1}{12} 
V^{ab} W^{cd} \bomega_{ac}{}^{e} \bomega_{bde}\nn\\&& -  \frac{1}{12} 
V^{ab} W^{cd} \bomega_{ca}{}^{e} \bomega_{dbe} -  \frac{1}{24} 
e^{\vp} \bH_{cde} V^{ab} V^{cd} \bomega^{e}{}_{ab} + 
\frac{1}{24} e^{-\vp}\bH_{cde} W^{ab} W^{cd} \bomega^{e}{}_{ab} \nn\\&&+ 
\frac{1}{12} e^{\vp} \bH_{bcd} V^{ab} \nabla_{a}V^{cd} -  
\frac{1}{12}e^{-\vp}\bH_{bcd} W^{ab} \nabla_{a}W^{cd}  + 
\frac{1}{24} e^{\vp} \bH_{bcd} V_{a}{}^{b} V^{cd} 
\nabla^{a}\vp \nn\\&&+ \frac{1}{24}e^{-\vp}\bH_{bcd} W_{a}{}^{b} W^{cd} 
\nabla^{a}\vp  + \frac{1}{12} V_{a}{}^{b} 
W^{cd} \bomega_{bcd} \nabla^{a}\vp + \frac{1}{12} V^{bc} 
W_{a}{}^{d} \bomega_{dbc} \nabla^{a}\vp \nn\\&&-  \frac{1}{12} 
W^{ab} \bomega_{a}{}^{cd} \nabla_{b}V_{cd} -  \frac{1}{12} 
W^{bc} \nabla^{a}\vp \nabla_{c}V_{ab} -  \frac{1}{12} V^{bc} 
\nabla^{a}\vp \nabla_{c}W_{ab} -  \frac{1}{6} V^{ab} 
W_{a}{}^{c} \nabla_{c}\nabla_{b}\vp\nn\\&& -  \frac{1}{6} V^{ab} 
\bomega_{a}{}^{cd} \nabla_{d}W_{bc} + \frac{1}{12} e^{\vp} 
\bH_{bcd} V^{ab} \nabla^{d}V_{a}{}^{c} -  \frac{1}{12} e^{-\vp}\bH_{bcd} W^{ab} 
\nabla^{d}W_{a}{}^{c}\Big]\labell{trans}
\eeqa
which is not invariant under the local Lorentz-transformation and  is  non zero.  However, the integrand might be cancelled by some total derivative terms or might be cancelled by some  terms at order $\alpha'$ that are produced by the transformation of the circular reduction of the leading order bulk action \reef{baction} under appropriate  deformations of the Buscher rules at order $\alpha'$.  Since the leading order action \reef{baction} is invariant under the local Lorentz-transformations,   the deformed Buscher rules and  the total derivative terms  should include terms that are not  invariant under the local Lorentz-transformations, \ie they should  include, among other fields, the spin connection $\bomega_{abc}$.

The circular reduction of  the leading order bulk action \reef{baction} is 
\beqa
S^{(0)}&=& -\frac{2}{\kappa'^2}\int d^{9}x\sqrt{-\bg}\, e^{-2\bphi}\Big[\bar{R}-\nabla^a\nabla_a\vp-\frac{1}{4}\nabla_a\vp \nabla^a\vp-\frac{1}{4}(e^{\vp}V^2 +e^{-\vp}W^2)\nn\\
&&\qquad\qquad\qquad\qquad\qquad+4\nabla_a\bphi\nabla^a \bphi+2\nabla_a\bphi\nabla^a\vp-\frac{1}{12}\bH_{abc}\bH^{abc}\Big]\labell{R}
\eeqa
which is invariant under the base space local Lorentz-transformations. The transformation of this action under the deformed Buscher rules \reef{T22} produces the following terms at order $\alpha'$:
\beqa
\alpha'\Delta (S^{(0)})&=& -\frac{2\alpha' }{\kappa'^2}\int d^{9}x \sqrt{-\bg} \,  e^{-2\bphi}\Big[-4 \Big(\frac{1}{2}\bar{R} +2\prt_c\bphi\prt^c\bphi -\frac{1}{8}\prt_c\vp\prt^c\vp-\frac{1}{24}\bH^2-\frac{1}{8}e^\vp V^2\nn\\
 &&-\frac{1}{8}e^{-\vp}W^2+\frac{1}{2}\nabla_c\nabla^c\vp-\prt_c\bphi\prt^c\vp\Big)\Delta\bphi+
\frac{1}{4}\Big(  e^\vp V^2-e^{-\vp}W^2\Big)\Delta\vp\nn\\
 &&+\frac{1}{2}e^{-\vp/2}\prt_b\vp W^{ab}\Delta g_a -\frac{1}{2}e^{\vp/2}\prt_b\vp V^{ab}\Delta b_a -\frac{1}{6}\bH^{abc}\Delta\bH_{abc}\nn\\
 &&+\frac{1}{2}(\prt_a\vp+4\prt_a\bphi)\nabla^a(\Delta\vp)-\nabla_a\nabla^a(\Delta\vp)-2(\prt_a\vp-4\prt_a\bphi)\nabla^a(\Delta\bphi)
 \nn\\
&&+e^{-\vp/2}W_{ab}\nabla^b(\Delta g^a)+e^{\vp/2}V_{ab}\nabla^b(\Delta b^a)\Big] 
\eeqa
Since the background has boundary, one has  to keep track of the total derivative terms. Hence, unlike in \cite{Garousi:2019wgz}, we do not use the integration by part to write the above transformations as the equations of motion multiplied by the deformations. The transformation \reef{trans} is odd-parity, hence, the deformations $\Delta\bphi$,  $\Delta\vp$, $\Delta b_a$ must include odd number of $\bH, W$  and  the deformations $\Delta g_a$,  $\Delta\bH_{abc}$ must include even number of $\bH, W$.  They have different parity with respect to the deformations \reef{dbH2}. 

The base space torsion $\bH$ satisfies the following  Bianchi identity \cite{Kaloper:1997ux}:
 \beqa
 \prt_{[a} \bH_{bcd]}&=&-\frac{3}{2}V_{[ab}W_{cd]}\labell{anB}
 \eeqa
This causes that the correction $\Delta \bH_{abc}$ to be  related to the corrections  $\Delta g_a$,  $\Delta b_a$ through the following relation:
\beqa
\Delta\bH_{abc}&=&\tilde H_{abc}-3e^{-\vp/2}W_{[ab}\Delta b_{c]}-3e^{\vp/2}V_{[ab}\Delta g_{c]}
\eeqa
where $\tilde H_{abc}=3\prt_{[a}\tB_{bc]}$  and $\tB_{ab}$  is a $U(1)\times U(1)$ gauge invariant 2-form constructed from the base space fields at order $\alpha'$ which is even under the parity. We find that there is no even-parity 2-form at order $\alpha'$ that are constructed from the base space fields, \ie   $\tilde H_{abc}=0$. 

Since the deformed Buscher rules must satisfy the $\MZ_2 $-algebra, the deformations at order $\alpha'$ must satisfy the following relations \cite{Garousi:2019wgz}:
\beqa
\Delta\vp-\Delta\vp|_{\vp\rightarrow -\vp,V\rightarrow W,W\rightarrow V}&=&0\nn\\
\Delta\bphi+\Delta\bphi|_{\vp\rightarrow -\vp,V\rightarrow W,W\rightarrow V}&=&0\nn\\
\Delta b+\Delta g|_{\vp\rightarrow -\vp,V\rightarrow W,W\rightarrow V}&=&0 \labell{ddddd}
\eeqa
One finds there is no odd-parity deformation $\Delta\bphi$ at order $\alpha'$ that satisfies the second relation above. Hence, $\Delta\bphi=0$. The assumption that the transformations at order $\alpha'$ should not include the second derivative terms, one finds the following terms for the deformations  $\Delta\vp, \Delta g$:
\beqa
\Delta\vp&=&e_1 V^{ab}W_{ab}+e_2\bH_{abc}\bomega^{abc}\labell{dpg}\\
\Delta g_a&=&b_1 e^{-\vp/2}\bH_{abc}W^{bc}+b_2 e^{\vp/2}\bomega_{abc}V^{bc}+b_3e^{\vp/2}\bomega_{bac}V^{bc}\nn\\&&
+b_4e^{\vp/2}\bomega^c{}_{bc}V_a{}^b+b_5 e^{\vp/2}V_{ab}\nabla^b\vp+b_6 e^{\vp/2}V_{ab}\nabla^b\bphi\nn
\eeqa
where $e_1,e_2,b_1,\cdots b_6$ are some constants. The deformation  $\Delta b$ is related to  $\Delta g$ by using the last relation in \reef{ddddd}. Note that one may include the term $e^{\vp/2}\nabla_b V_a{}^b$ to the second  deformation above which is at order $\alpha'$ and is also  even-parity. However, this term changes the data on the boundary at order $\alpha'$. Hence, this term breaks the $O(1,1)$ symmetry at the boundary. We will comment on this point in the Discussion section.

We  include all possible total covariant derivative terms into our calculations. The most general  total derivative terms  are the following:
\beqa
\cJ^{(1)}&=&-\frac{2\alpha'}{\kappa'^2}\int d^{9} x\sqrt{-\bg}\, \nabla_{a}\Bigg[e^{-2\bphi}I^{(1)a}\Bigg]\labell{J1}
\eeqa
where the vector $I^{(1)a}$ is all contractions of the fluxes $\bomega,\bH, V,W,\nabla\bphi, \nabla\vp$ and their covariant derivatives at three-derivative order  which are odd-parity.  

If the action  \reef{trans} is going to be  invariant under the deformed Buscher rules, then the following  relation must be satisfied:
\beqa
\delta S^{(1)}_O+\alpha'\Delta(S^{(0)})+\cJ^{(1)}&=&0\labell{cons1}
\eeqa
To  check this relation explicitly, one  must include in it all  the Bianchi identities. To impose them, we write the curvatures, the spin connection and the covariant derivatives in \reef{cons1} in terms of partial derivatives and frame $\bar{e}_a{}^\ti$, and write the field strengths $\bH,V,W$ in terms of potentials $\bb_{ab},b_a,g_a$. In this way all the Bianchi identities satisfy automatically. Then the equation \reef{cons1} can be written in terms of  non-covariant  independent terms. If the above relation is correct, then the coefficients of the independent terms should be zero, \ie there should be total derivative terms and appropriate corrections for the Buscher rules  that  make the above relation to be satisfied. 

We have found that using the following current in the total derivative terms: 
\beqa
I^{(1)a}&=&\frac{1}{3} W^{bc}\nabla_c V^a{}_b+\frac{1}{3}V^{bc}\nabla_{c}W^a{}_b\labell{I1}
\eeqa
and using the following deformations:
\beqa
\Delta\vp&=&-\frac{1}{6} V^{ab}W_{ab}\labell{dpg1}\\
\Delta g_a&=&-\frac{1}{24} e^{-\vp/2}\bH_{abc}W^{bc}+\frac{1}{12} e^{\vp/2}\bomega_{abc}V^{bc}+\frac{1}{12} e^{\vp/2}V_{ab}\nabla^b\vp\nn
\eeqa
 then all terms on the left-hand side of  \reef{cons1}  are cancelled, \ie the relation \reef{cons1} is $exactly$ satisfied. Similar calculations as above have been done in \cite{Garousi:2023diq} to check if the coupling \reef{CS} is invariant under the $O(D,D)$-transformations without using the KK reduction.   In that case, one finds there is $one$  term in the transformation of the action \reef{CS} under the non-geometrical subgroup of  $O(D,D)$-group that can  not be cancelled by any total derivative term nor by any deformation of the non-geometrical transformations.
 
Since there are residual  total derivative terms in \reef{I1}, the above result indicates so far that the coupling \reef{CS} is invariant under the $O(1,1)$-group for the closed spacetime manifolds. However, since the $O(1,1)$  is a global symmetry, one expects from the background independence assumption that the effective action should have the $O(1,1)$ symmetry even for the open manifolds that have boundary. In that cases,  the total derivative terms must  be  cancelled by the transformation of the circular reduction of the boundary term in the leading order effective action \reef{baction}. 
 
 The circular reduction of  the  boundary term in  \reef{baction} is 
 \beqa
 \prt S^{(0)}&=&-\frac{4}{\kappa'^2}\int d^{8}\sigma\,\sqrt{| \tg|}\, e^{-2\bphi}\Big[  \bg^{ab}\bK_{ab}+\frac{1}{2} n^a\nabla_a\vp\Big]\labell{redcebs}
\eeqa
 where $\tg$ is determinant of the base space induced metric $\tg_{\ta\tb}=\frac{\prt x^a}{\prt\sigma^\ta}\frac{\prt x^b}{\prt\sigma^\tb}\bg_{ab}$.
 Since, the base space metric $\bg$ and dilaton $\bphi$ are  invariant under the deformed Buscher rules, the transformation of the above action under the deformation \reef{T22} is
 \beqa
 \alpha'\Delta (\prt S^{(0)})&=&-\frac{2\alpha'}{\kappa'^2}\int d^{8}\sigma\,\sqrt{| \tg|}\, e^{-2\bphi}\Big[ n^a\nabla_a(\Delta\vp)\Big]\nn\\
 &=&\frac{\alpha'}{3\kappa'^2}\int d^{8}\sigma\,\sqrt{| \tg|}\, e^{-2\bphi}\Big[ n^a\nabla_a(V^{bc}W_{bc})\Big]\labell{dpS}
\eeqa
 where in the second line we have replaced the deformation found in \reef{dpg1}. 
 
 On the other hand, inserting the current \reef{I1} into \reef{J1}, and using the Stokes' theorem, one finds the following boundary terms in the base space:
 \beqa
\cJ^{(1)}&=&-\frac{2\alpha'}{3\kappa'^2}\int d^{8} \sigma\sqrt{|\tg|}\, e^{-2\bphi} n_{a}\Bigg[ W^{bc}\nabla_c V^a{}_b+V^{bc}\nabla_{c}W^a{}_b\Bigg]\nn\\&=&
-\frac{\alpha'}{3\kappa'^2}\int d^{8} \sigma\sqrt{|\tg|}\, e^{-2\bphi} n_{a}\Bigg[ W^{bc}\nabla_a V_{bc}+V^{bc}\nabla_{a}W_{bc}\Bigg]
\labell{J2}
\eeqa
where in the second line we have used the Bianchi identities $\nabla_{[a}V_{bc]}=0=\nabla_{[a}W_{bc]}$. The above residual boundary term is $exactly$ cancelled by the transformation of the leading order boundary term  \reef{dpS}. 

\section{Discussion}

In this paper we have shown that the parity odd coupling at order $\alpha'$  in the heterotic string theory which is produced by the replacement of 
$H$ in \reef{baction} with the new field strength \reef{replace}, is invariant under $O(1,1)$-transformations after reducing the coupling on a circle.  This calculation for the closed spacetime manifolds   has been done in \cite{Garousi:2019wgz}. In the present  paper, we extend the calculation to the open spacetime manifolds that  have boundary. This calculation confirms that the $O(1,1)$  symmetry is in fact the symmetry of the combination of  the bulk and boundary actions, \ie
\beqa
S+\prt S&\rightarrow S+\prt S
\eeqa
where the  actions and the non-geometrical $\MZ_2$-transformations have $\alpha'$-expansions. The above symmetry also confirms  the proposal that the effective actions at the critical dimension are background independent, \ie the $O(1,1)$ is symmetry of the closed and  the open spacetime manifolds. 
 
The above calculations also confirm the assumption that in the least action principle,  the data on the boundary for  the effective action at order $\alpha'$ are the values of the massless fields and their first derivatives \cite{Garousi:2021cfc}, \ie   the values of the second derivatives of the massless fields   are not known on the boundary for the effective actions at order $\alpha'$. The T-duality  transformations at order $\alpha'$ should respect these data.  This dictates that the T-duality  transformations  should not involve the second derivative of the massless  fields because such transformations when applied on the data at the leading order $\alpha'$, would transform the  values of the massless fields at the leading order of $\alpha'$ to the  values of the second derivatives of the massless fields at order $\alpha'$ which are not known for the effective action at order $\alpha'$. In other words,  such transformations would  break the $O(1,1)$  symmetry at the boundary. As a result,  there would be no  boundary term that is consistent with  this symmetry.  In fact, we have considered  the allowed transformations in \reef{dpg} and found the consistent result that there is no parity odd boundary term at order $\alpha'$. 

If we had included the second derivative term $e^{\vp/2}\nabla_b V_a{}^b$ into $\Delta g_a$ in \reef{dpg} and insisted that its coefficient is non-zero, then we would find that the bulk relation \reef{cons1} is satisfied provided that one uses the following total derivative terms:
\beqa
I^{(1)a}&=&-\frac{1}{24}e^{\vp}\bH_{bcd}V^{ab}V^{cd}+\frac{1}{24}e^{-\vp}\bH_{bcd}W^{ab}W^{cd}+\frac{1}{12}V_b{}^c W^a{}_c\nabla^b\vp+\frac{1}{12}V^{ac}W_{bc}\nabla^b\vp\nn\\&&+
\frac{1}{4} W^{bc}\nabla_c V^a{}_b+\frac{5}{12}V^{bc}\nabla_{c}W^a{}_b\labell{I2}
\eeqa
and the following corrections to the Buscher rules:
\beqa
\Delta\vp&=&-\frac{1}{6} V^{ab}W_{ab}\labell{def}\\
\Delta g_a&=&-\frac{1}{12} e^{\vp/2}\nabla_b V_a{}^b+\frac{1}{12} e^{\vp/2}\bomega_{abc}V^{bc}+\frac{1}{6} e^{\vp/2}V_{ab}\nabla^b\bphi\nn
\eeqa
Note that the first deformation above is the same as the deformation in \reef{dpg1}. Using the above deformations, one finds the same transformation of the leading order boundary term as \reef{dpS}. However,  this term would not be cancelled by the total derivative term \reef{I2} any more. Moreover, there is no parity odd boundary coupling at order $\alpha'$ to cancels the remaining term. This indicates that the  $O(1,1)$-symmetry would be  broken at the boundary, as expected from choosing the wrong transformation that breaks the data on the boundary.

It has been proposed in \cite{Garousi:2021yyd} that  the field redefinitions at order $\alpha'$ which do not change the data on the boundary should be the restricted field redefinitions, \ie    the metric has no transformation and all other fields have transformations that  involve only the first derivative of the massless fields.  To clarify this point, we  consider the leading order action \reef{R}. Up to a total derivative term, this action  has the second derivative on metric $\bg$ and the first derivative on all other base space fields, collectively called $\psi$. Hence, under the  field redefinitions at order $\alpha'$, \ie $\bg\rightarrow \bg +\alpha'\delta\bg$ and $\psi\rightarrow \psi+\alpha'\delta\psi$, the second derivative of $\delta\bg$ and the first derivative of $\delta\psi$ appear in the bulk. Using the Stokes' theorem, $\delta\psi$ itself and the first derivative of $\delta\bg$   appear on the boundary. On the other hand, the field redefinitions should transform  the data at the leading order of $\alpha'$ which are the values of the massless fields, to the data at order $\alpha'$ which are the values of the massless fields and their first derivatives. If $\delta\bg$ includes the  first derivative of the base space fields, then the first derivative of $\delta\bg$ produces the second derivative of the base space fields which are not known on the boundary for the effective action at order $\alpha'$. Such field redefinitions then ruin the data on the boundary. Hence, the field redefinitions at order $\alpha'$ which do not change the data on the boundary is that $\delta\bg$ should not include the first derivative of the base space  fields, and all other fields should have transformations that  involve only the first derivative of the massless fields \cite{Garousi:2021yyd}. Note that the data on the boundary at order $\alpha'$ include the values of the metric and its first derivative, however, the allowed field redefinitions do not change the metric. 

The field redefinition of the metric, in particular,  implies that the length of the unite vector of the boundary is invariant under the field redefinitions. We expect this  property should be satisfied for the field redefinitions at higher orders of $\alpha'$ as well, \ie the metric is not changed under the field redefinitions at any order of $\alpha'$. This is consistent with the proposal that the data for the effective action at order $\alpha'^n$ is the values of the massless fields and their derivatives up to order $n$  \cite{Garousi:2021cfc}. To clarify it we use the iterative method. At order $\alpha'^2$,  if  the field redefinition $\delta\bg$ includes the second derivative of the base space fields, then the first derivative of $\delta\bg$ which appears on the boundary, produces the third  derivative of the base space fields which are not known on the boundary for the effective action at order $\alpha'^2$. Such field redefinitions then ruin the data on the boundary at order $\alpha'^2$. Hence, $\delta\bg$ should not include the second derivative of the base space  fields either.  At order $\alpha'^3$, if $\delta\bg$ includes the third  derivative of the base space fields, then the first derivative of $\delta\bg$ produces the fourth  derivative of the base space fields which are not known on the boundary for the effective action at order $\alpha'^3$. Such field redefinitions then ruin the data on the boundary at order $\alpha'^3$. Hence, $\delta\bg$ should not include the third  derivative of the base space  fields. Similarly for the higher orders of $\alpha'$. Hence, the invariance of the above  data    under the field redefinitions requires the metric to be invariant under the field redefinitions.

We have seen that the study of the invariance of the circular reduction of the coupling \reef{CS} under the non-geometrical subgroup of $O(1,1)$ confirms that the data on the boundary for the effective action at order $\alpha'$ are the values of the massless fields and their first derivatives. The replacement \reef{replace} into the leading order action \reef{baction} produces also  $one$  bulk term at order $\alpha'^2$ and no boundary term, \ie $\Omega^2$. This term is even under the parity and is not invariant under the local Lorentz-transformations. There is no other even-parity bulk couplings at this order in the heterotic string theory. The circular reduction of this term should also  be invariant under appropriate  deformed Buscher rules at order $\alpha'^2$. From this study one may find   the data on the boundary and compare them with the data proposed in \cite{Garousi:2021cfc}  for the effective action at order $\alpha'^2$. We leave the details of this calculations for the future works. 


\end{document}